\documentclass{ws-mpla}

\begin{document}

\markboth{H. Bartko for the MAGIC collaboration}
{Observation of Galactic Sources of Very High Energy $\gamma$-Rays with the MAGIC Telescope}

\catchline{}{}{}{}{}

\title{Observation of Galactic Sources of Very High Energy $\gamma$-Rays with the MAGIC Telescope}

\author{\footnotesize H. BARTKO for the MAGIC collaboration}

\address{MPI f\"ur Physik, Werner-Heisenberg-Institut,\\
F\"ohringer Ring 6, D-80805 M\"unchen, Germany\\
E-mail: hbartko@mppmu.mpg.de}

\maketitle

\pub{Received (Day Month Year)}{Revised (Day Month Year)}

\begin{abstract}
The MAGIC telescope with its 17m diameter mirror is today 
the largest operating single-dish Imaging Air Cherenkov Telescope (IACT).
It is located on the Canary Island La Palma, at an altitude of
2200m above sea level, as part of the Roque de los
Muchachos European Northern Observatory. The MAGIC telescope detects celestial very high energy $\gamma$-radiation in the energy band between about 50~GeV and 10~TeV. Since the autumn of 2004 MAGIC has been taking data 
routinely, observing various
objects, like supernova remnants (SNRs), $\gamma$-ray binaries, Pulsars, Active Galactic Nuclei (AGN) and Gamma-ray Bursts (GRB).
We briefly describe the observational strategy, the procedure implemented for the data analysis, and discuss the results of observations of Galactic Sources.
\end{abstract}

\ccode{PACS Nos.: 95.85.Pw, 95.85.Ry}

\section{Introduction -- The MAGIC Telescope} 
\label{sec:intro}
MAGIC \cite{MAGIC-commissioning,CortinaICRC} 
is currently the largest single dish Imaging
Air Cherenkov Telescope (IACT) in operation. Located on the Canary
Island La Palma ($28.8^\circ$N, $17.8^\circ$W, 2200~m a.s.l.), it 
has a 17-m diameter tessellated parabolic mirror,
supported by a light weight carbon fiber frame. It is equipped
with a high quantum efficiency 576-pixel $3.5^\circ$ field-of-view photomultiplier
camera. The analog signals are transported via optical fibers to
the trigger electronics
and the 2 GSamples/s FADC system \cite{MUX_tests}. 


The MAGIC telescope can operate under moderate moonlight or twilight conditions \cite{MAGIC_moon}. For these conditions, no change in the high voltage settings is necessary as the camera PMTs were especially designed to avoid high currents.

The physics program of the MAGIC telescope includes both,
topics of fundamental physics and astrophysics. In this paper 
the observations of galactic sources are reviewed. 
The observations of extragalactic sources are presented
elsewhere, see e.g.~\cite{errando}.

\section{Data Analysis}
\label{sec:data_analysis}

The data analysis is generally carried out using the standard MAGIC
analysis and reconstruction software \cite{Magic-software}, the
first step of which involves the FADC signal reconstruction and the 
calibration of the raw data
\cite{MAGIC_calibration,MAGIC_signal_reco}. After calibration, image-cleaning
tail cuts are applied (see e.g. \cite{Fegan1997}).
The camera images are parameterized by
image parameters \cite{Hillas_parameters}. The
Random Forest method (see \cite{RF,Breiman2001} for a detailed
description) was applied for the $\gamma$/hadron separation (for a
review see e.g. \cite{Fegan1997}) and the energy estimation.



For each event, the arrival direction of the primary $\gamma$-ray
candidate in sky coordinates is estimated using the DISP-method resulting in VHE $\gamma$-ray sky maps
\cite{wobble,Lessard2001,MAGIC_disp}. The angular resolution of this procedure is $\sim 0.1^\circ$, while the source localization in the sky is provided with a systematic error of $1'$ \cite{MAGIC_Crab}.

The differential
VHE $\gamma$-ray spectrum
($\mathrm{dN}_{\gamma}/(\mathrm{dE}_{\gamma} \mathrm{dA}
\mathrm{dt})$ vs. true $\mathrm{E}_{\gamma}$) is 
corrected (unfolded) for the instrumental energy resolution
\cite{Anykeev1991,Bertero1989,MAGIC_unfolding}. 
All fits to the spectral points take into account the correlations between the spectral points that are introduced by the unfolding procedure.
%
%

%
The systematic error in the flux level determination depends on the slope of the $\gamma$-ray spectrum. It is
typically estimated to be 35\% and the systematic error in the
the spectral index is 0.2 \cite{MAGIC_GC,MAGIC_Crab}.


\section{Galactic Sources} \label{sec:galactic}

The observations with the MAGIC telescope 
included the following
types of objects: VHE $\gamma$-ray sources coincident with supernova remnants (section \ref{sec:snr}), the Galactic Center (section \ref{sec:gc}), the $\gamma$-ray binary LS~I~+61~303 (section \ref{sec:lsi}), pulsars and pulsar wind nebulae (section \ref{sec:crab}).


\vspace{-0.3cm}
\subsection{Unidentified VHE $\gamma$-ray sources coincident with Supernova remnants}
\label{sec:snr}


Shocks produced by supernova explosions are assumed to be the source
of the galactic component of the cosmic ray flux~\cite{zwicky}. 
In inelastic collisions of high energy cosmic rays with ambient matter $\gamma$-rays and neutrinos are produced. These neutral particles give direct information about their source, as their trajectories are not affected by the Galactic and extra Galactic magnetic fields in contrast to the charged cosmic rays. However, not all VHE $\gamma$-rays from galactic sources are due to the interactions of cosmic rays with ambient matter. There are also other mechanisms for the production of VHE $\gamma$-rays like the inverse Compton up-scattering of ambient low energy photons by VHE electrons. For each individual source of VHE $\gamma$-rays, the physical processes of particle acceleration and $\gamma$-ray emission in this source have to be determined. A powerful tool is the modeling of the multiwavelength emission of the source taking into account the ambient gas density as traced by CO observations \cite{Torres2002}.

Within its program of observation of galactic sources, MAGIC has
taken data on a number of supernova remnants, resulting in the discovery of VHE $\gamma$-ray emission from a source in the SNR IC443, MAGIC J0616+225 \cite{MAGIC_IC443}. Moreover, two recently discovered VHE $\gamma$-ray sources, which are spatially coincident with SNRs, HESS~J1813-178 and HESS~J1834-087 \cite{Aharonian2005b} have been observed with the MAGIC telescope \cite{MAGIC_1813,MAGIC_1834}. 

\begin{figure}[h]
  \begin{minipage}[t]{0.47\textwidth}
  \centering
  \includegraphics[width=60mm]{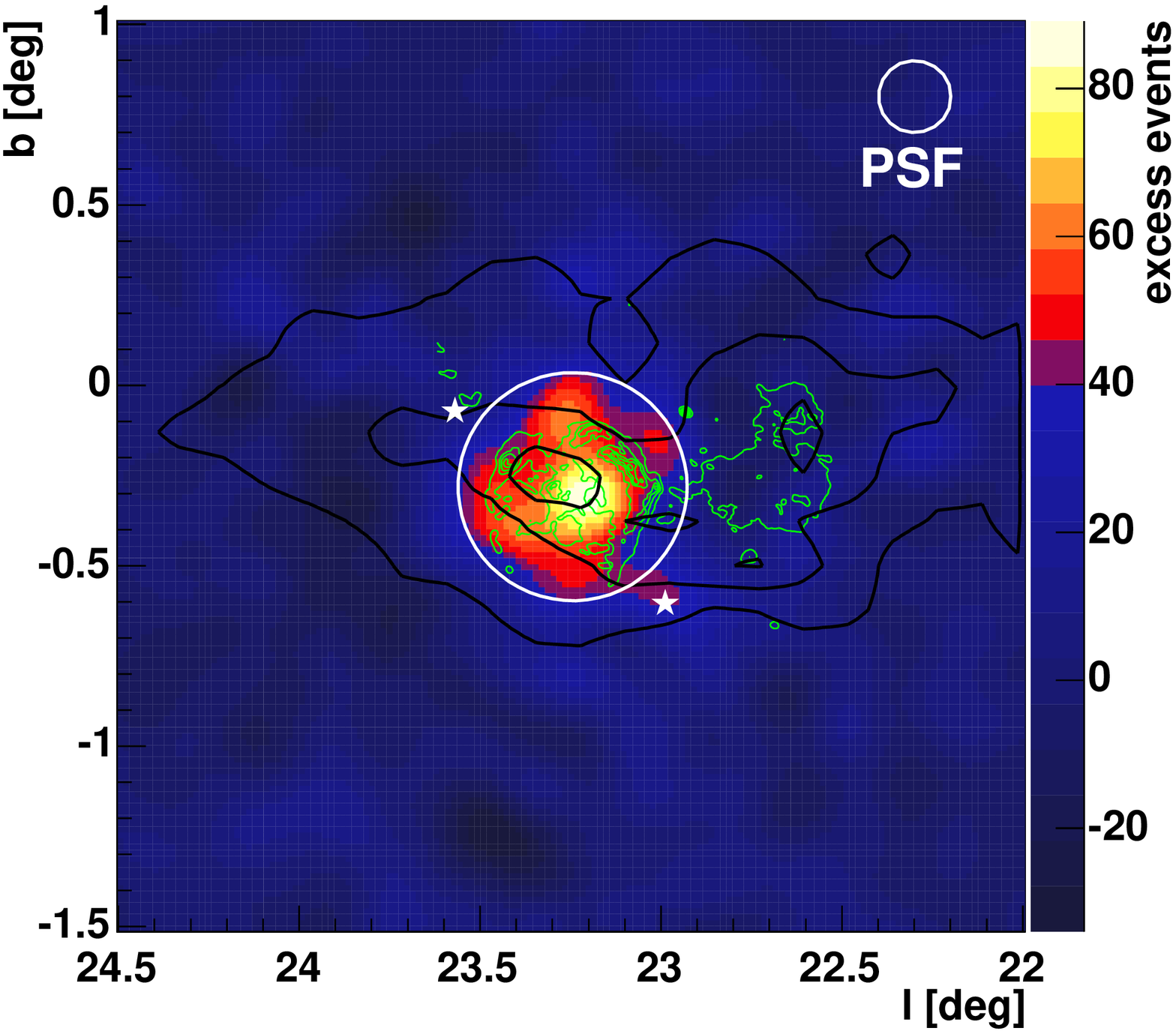}
  \vspace{-5mm}
  \caption{{\small Sky map
 of $\gamma$-ray candidate events (background
subtracted) in the direction of HESS J1834-087 for an energy threshold
of about 250~GeV. The source is clearly extended with respect to
the MAGIC PSF (small white circle). The two white stars denote the tracking positions of the MAGIC telescope. Overlayed are $^{12}$CO  emission contours (black) from $^{28}$ and contours of 90 cm VLA radio data
from $^{49}$ (green). The $^{12}$CO contours are at
25/50/75 K km/s, integrated from 70 to 85 km/s in velocity, the
range that best defines the molecular cloud associated with W41. The contours of the radio emission are at
0.04/0.19/0.34/0.49/0.64/0.79 Jy/beam, chosen for best showing
both SNRs G22.7-0.2 and G23.3-0.3 at the same time. Clearly, there
is no superposition with SNR G22.7-0.2. The central white circle
denotes the source region integrated for the spectral analysis. $^{6}$.} 
\label{fig:sky_W41} }

  \end{minipage}%
	\hspace{0.05\textwidth}
  \begin{minipage}[t]{0.47\textwidth}
  \centering
  \includegraphics[width=63mm]{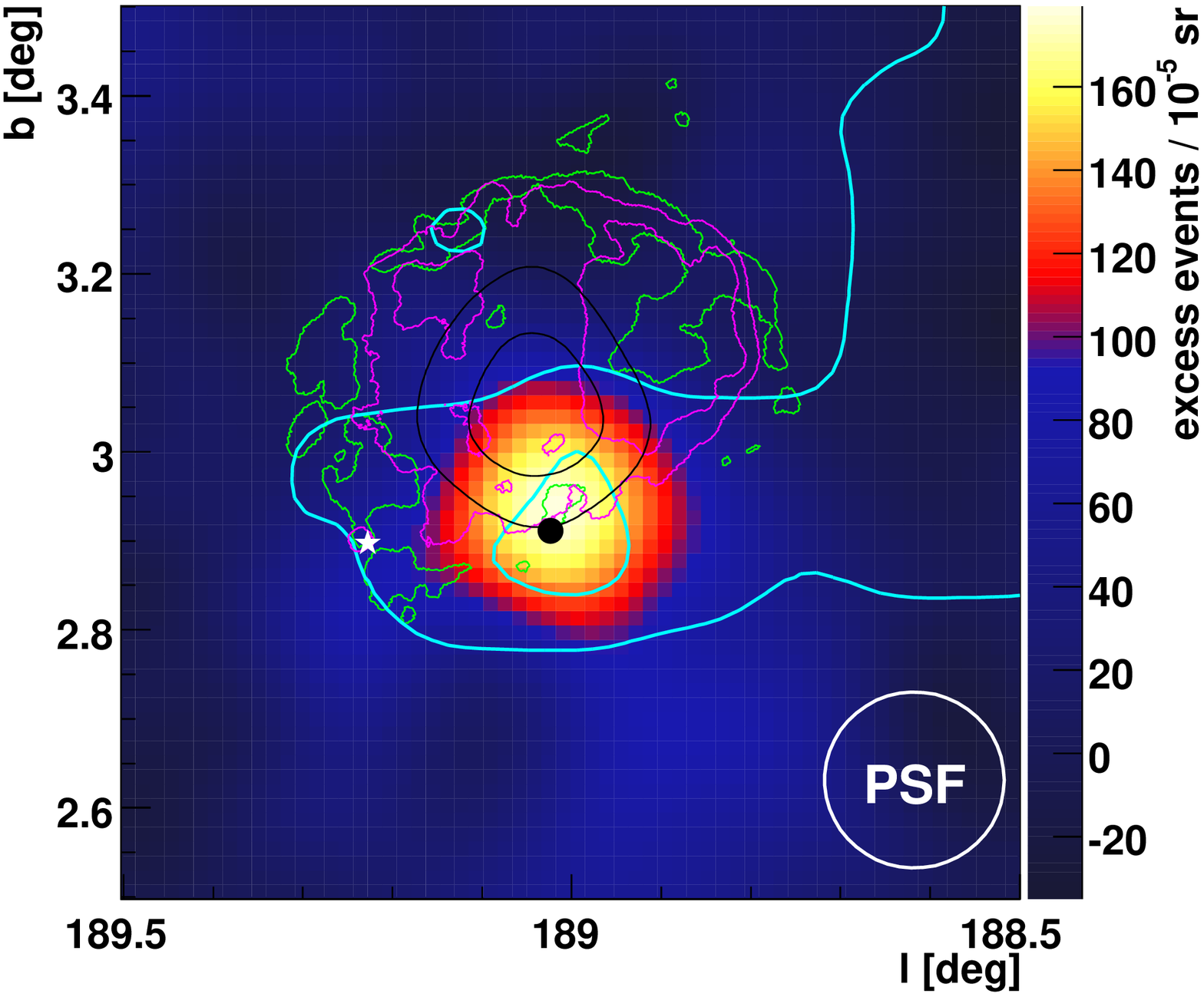}
  \vspace{-5mm}
  \caption{{\small Sky map of $\gamma$-ray candidate events (background
subtracted) in the direction of MAGIC~J0616+225 for an energy threshold
of about 150~GeV. The cyan $^{12}$CO contours $^{28}$ are at
7 and 14 K km/s, integrated from -20 to 20 km/s in velocity, the
range that best defines the molecular cloud associated with IC~443.
The  green contours of 20 cm VLA radio data
$^{25}$ are at
5 mJy/beam, chosen for best showing
both the SNR IC~443. 
The purple Rosat X-ray contours $^{14}$ are at 700 and 1200 counts / $6 \cdot 10^{-7}$ sr. The black EGRET contours $^{36}$ represent a 68\% and 95\% statistical probability that a single source lies within the given contour.
The white star denotes the position of the pulsar CXOU J061705.3+222127 $^{45}$. The black dot shows the position of the 1720 MHz OH maser $^{24}$. The white circle shows the MAGIC PSF of $\sigma = 0.1^{\circ}$. $^{10}$.}  \label{fig:sky_IC443}}
\end{minipage}%

\end{figure}

\textbf{IC443} is a well-studied shell-type SNR near the Galactic Plane with a diameter of 45' at a distance of about 1.5~kpc. It is a prominent source and it has been studied from radio waves to $\gamma$-rays of energies around 1~GeV. \cite{Gaisser1998,Baring1999} and \cite{Kaul2001} extrapolated the energy spectrum of 3EG~J0617+2238 into the VHE $\gamma$-ray range and predicted readily observable fluxes. Nevertheless, previous generation IACTs have only reported upper limits to the VHE $\gamma$-ray emission \cite{Khelifi2003,Holder2005}.
The observation of IC~443 using the MAGIC Telescope has
led to the discovery of a new source of VHE $\gamma$-rays, MAGIC~J0616+225. The flux level of MAGIC~J0616+225 is lower and the energy spectrum (fitted with a power law of slope $\Gamma=-3.1\pm 0.3$) is softer than the predictions.
The coincidence
of the VHE $\gamma$-ray source with SNR IC~443 suggests this
SNR as a natural counterpart. 
A massive molecular cloud and OH maser emissions are located at
the same sky position as that of MAGIC~J0616+225, see figure \ref{fig:sky_IC443}. This suggests that a hadronic origin of the VHE $\gamma$-rays is possible. 
However, other mechanisms for the VHE $\gamma$-ray emission cannot be excluded yet.

\textbf{HESS~J1834-087} is spatially coincident with the SNR G23.3-0.3 (W41). W41 is an asymmetric shell-type SNR, with a diameter of 27' at a distance of $\sim 5$~kpc. It is a prominent radio source, and only recently \cite{Landi2006} found a faint X-ray source within the area of W41 in data from the Swift satellite and \cite{ti06} found an extended X-ray feature spatially coincident with the VHE $\gamma$-ray emission. As in the case of IC~443, the VHE $\gamma$-radiation of W41 is associated with a large molecular complex called "[23,78]''~\cite{da86}, see figure \ref{fig:sky_W41}. Although the mechanism responsible
for the VHE $\gamma$-radiation has not yet been clearly identified,
it could be produced by high energy hadrons interacting with the
molecular cloud.


\textbf{HESS~J1813-178} is spatially coincident with SNR G12.8-0.0 with a diameter of 2' at a distance of $\sim 4$~kpc. It exhibits relatively faint radio and X-radiation. This source is also located in a relatively high-density environment \cite{Lemiere2005}. \cite{he07} discovered a putative pulsar wind nebula and associated it with the VHE $\gamma$-ray source.


These results confirm that Galactic VHE $\gamma$-ray sources are usually spatially correlated with SNRs. Nevertheless, the exact nature of the parent particles of the VHE $\gamma$-rays, their acceleration (in SNR shocks or PWN), and the processes of $\gamma$-ray emission need (or still require) further study.

\subsection{The Galactic Center}
\label{sec:gc}

The Galactic Center region contains many remarkable objects which may be responsible for high-energy processes generating $\gamma$-rays: A super-massive black hole, supernova remnants, candidate pulsar wind nebulae, a high density of cosmic rays, hot gas and large magnetic fields. Moreover, the Galactic Center may appear as the brightest VHE $\gamma$-ray source from the annihilation of possible dark matter particles \cite{DM_ICRC} of all proposed dark matter particle annihilation sources. 


The Galactic Center was observed with the MAGIC telescope~\cite{MAGIC_GC} under large zenith angles, resulting in the measurement of a differential $\gamma$-ray flux, consistent with a steady, hard-slope power law between 500~GeV and about 20~TeV, with a spectral index of $\Gamma=-2.2\pm 0.2$.
This result confirms the previous measurements by the HESS collaboration. The VHE $\gamma$-ray emission does not show any significant time variability; the MAGIC measurements rather affirm a steady emission of $\gamma$-rays from the GC region on time scales of up to one year.

The VHE $\gamma$-ray source is centered at (RA, Dec)=(17$^{\mathrm{h}}45^{\mathrm{m}}20^{\mathrm{s}}$, -29$^\circ2'$). The excess is point-like, its location is consistent with SgrA$^*$, the candidate PWN G359.95-0.04 as well as SgrA East.
The nature of the source of the VHE $\gamma$-rays has not yet been (or yet to be) identified. The power law spectrum up to about 20~TeV disfavours dark matter annihilation as the main origin of the detected flux.
The absence of flux variation indicates that the VHE $\gamma$-rays are rather produced in a steady object such as a SNR or a PWN, and not in the central black hole.



\subsection{The $\gamma$-ray binary LS~I~+61~303} 
\label{sec:lsi}

\begin{figure}[!t]
\centering
\includegraphics[width=\textwidth]{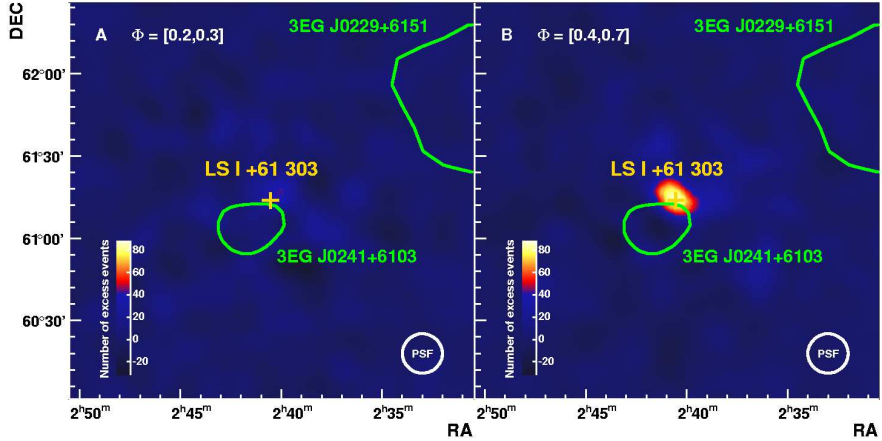}
\caption{\small Smoothed maps of $\gamma$-ray excess events above 400~GeV around LS~I~+61~303. (A) 15.5 hours corresponding to data around periastron, i.e. between orbital phases 0.2 and 0.3. (B) 10.7 hours at orbital phase between 0.4 and 0.7. The number of events is normalized in both cases to 10.7 hours of observation. The position of the optical source LSI +61 303 (yellow cross) and the 95\% confidence level contours for 3EG J0229+6151 and 3EG J0241+6103 (green contours) $^{36}$, are also shown. The bottom-right circle shows the size of the point spread function of MAGIC (1$\sigma$ radius). No significant excess in the number of $\gamma$-ray events is detected around periastron passage, while it shows up clearly (9.4$\sigma$ statistical significance) at later orbital phases, in the location of LS~I~+61~303. $^{5}$.
}
\label{fig:lsi-skymap}
\end{figure}

This $\gamma$-ray binary system is composed of a B0 main sequence star
with a circumstellar disc, i.e. a Be star, located at a distance of

$\sim$2 kpc. A compact object of unknown nature (neutron star or black
hole) is orbiting around it, in a highly eccentric ($e=0.72\pm0.15$)
orbit.
  
LS~I~+61~303 was observed with MAGIC for 54 hours
between October 2005 and March 2006~\cite{MAGIC_LSI}. 
The reconstructed $\gamma$-ray sky map is shown in
figure~\ref{fig:lsi-skymap}. The data were first divided into 
two different samples, around periastron passage (0.2-0.3)
and at higher (0.4-0.7) orbital phases. No significant excess in the
number of $\gamma$-ray events is detected around periastron passage,
whereas there is a clear detection (9.4$\sigma$ statistical significance) at
later orbital phases.
Two different scenarios were discussed to explain this high energy
emissions: the microquasar scenario where the $\gamma$-rays are produced
in a radio-emitting jet; or the pulsar binary scenario, where 
they are produced in the shock which is generated by the interaction
of a pulsar wind and the wind of the massive companion.

\subsection{Pulsars and Pulsar Wind Nebulae} \label{sec:crab}

\begin{figure}[!t]
\centering
\includegraphics[height=5cm]{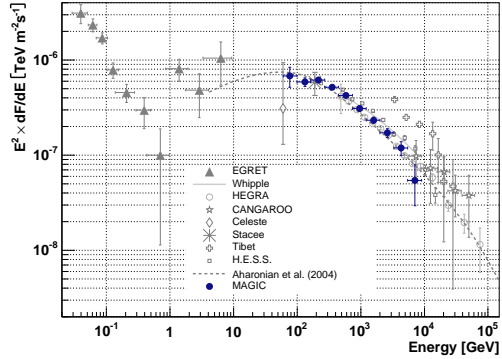}
\caption{{\small Spectral energy distribution of the  $\gamma$-ray emission of the Crab Nebula. The measurements below 10~GeV are from the EGRET, the measurements above are from ground-based experiments. Above 400~GeV the MAGIC data are in agreement with measurements of other IACTs. The dashed line represents a model prediction by $^{1}$. $^{11}$.}
}
\label{fig:crab}
\end{figure}

\begin{figure}[!t]
\centering
\includegraphics[height=5cm]{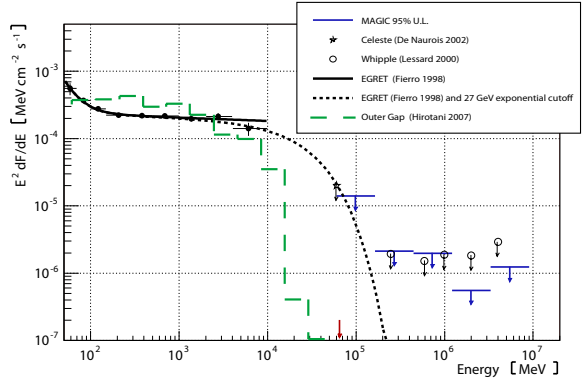}
\caption{{\small Upper limits (95\% CL.) on the pulsed $\gamma$-ray flux from the Crab Pulsar; upper limits in bins of energy are given by the blue points. The upper limit on the cutoff energy of the pulsed emission is indicated by the dashed line. The analysis threshold to derive the upper limit on the cutoff energy is indicated by the red arrow. $^{11}$.}
}
\label{fig:crab_pulsar}
\end{figure}

The Crab Nebula is a bright and steady emitter of GeV and TeV energies, 
and is therefore an excellent calibration candle. This object has
been observed intensively in the past, over a wide range of
wavelengths. 

The energy domain between 10 and 100~GeV is of particular interest, as both
the Inverse Compton peak of the spectral energy distribution 
and the cut-off of the pulsed emission is expected in this energy range.


A significant amount of MAGIC's observation time has been devoted to observing the Crab Nebula, both for
technical (because it is a strong and steady emitter) and astrophysical studies. A sample of
16~hours of selected data has been used to measure
the energy spectrum between 60~GeV and 9~TeV, and the result is shown in
figure~\ref{fig:crab}~\cite{MAGIC_Crab}. Also, a search
for pulsed $\gamma$-ray emission from the Crab Pulsar has been carried out. Figure~\ref{fig:crab_pulsar} shows the derived (95\% CL.) upper limits.



Pulsed $\gamma$-ray emission was also searched for from the pulsar PSR B1951+32. 
A 95\% CL. of $4.3 \cdot 10^{-11}$~cm$^-2$~sec$^{-1}$ was obtained for the flux of pulsed $\gamma$-ray emission for $E_{\gamma}>75$~GeV and of $1.5 \cdot 10^{-11}$~cm$^-2$~sec$^{-1}$ for the steady emission for $E_{\gamma}>140$~GeV \cite{MAGIC_1951}.

\section*{Acknowledgments}
We thank the IAC for the excellent working conditions at the
ORM in La Palma. The support of the
German BMBF and MPG, the Italian INFN, the Spanish CICYT is gratefully
acknowledged. This work was also supported by ETH research grant
TH-34/04-3, and the Polish MNiI grant 1P03D01028. 


\end{document}